\documentclass[amsmath,amssymb,aps,10pt,prd,letterpaper,nofootinbib,balancelastpage,notitlepage,superscriptaddress,twocolumn,floatfix,preprintnumbers]{revtex4-2}
\usepackage{graphicx}	% Standard figures 
\usepackage{ragged2e}	% for justified text in JHEP.bst
\usepackage{bm}		    % Bold math
\usepackage[sort&compress]{natbib}	 	% Standard reference package
	\setcitestyle{square,numbers,comma}	% Set citation style
\usepackage[colorlinks=true,urlcolor=blue,linkcolor=black,citecolor=red]{hyperref}	
\usepackage{slashed}
\renewcommand{\eqref}[2][]{Eq{#1}.~(\ref{eq:#2})}		% Equation reference
\newcommand{\orcid}[1]{\href{https://orcid.org/#1}{\,\includegraphics[width=8px]{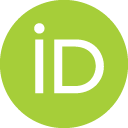}}}
\def\mdp{m_{\gamma'}}
\def\rhodm{\rho_\text{DM}}

\def\fov{\Omega_\text{FoV}}
\def\me{m_e}

\usepackage{color}
   \definecolor{k}{RGB}{0,0,0}
   \definecolor{r}{RGB}{255,0,0}
   \definecolor{R}{RGB}{255,0,0}
   \definecolor{o}{RGB}{255,94,1}
   \definecolor{O}{RGB}{255,165,0}
   \definecolor{p}{RGB}{148,55,255}
   \definecolor{P}{RGB}{148,55,255}
   \definecolor{g}{RGB}{0,150,50}
   \definecolor{G}{RGB}{0,150,50}
   \definecolor{b}{RGB}{0,0,255}
   \definecolor{B}{RGB}{0,0,255}

\begin{document}
%%%%%%%%%%%%%%%%%%%%%%%%%%%%%%%%%%%%%%%%%%%%%%%%%%%%%%%%%%%%%%%%%%%%%%%%%%%%%%%%%%%%%%%%%%
%%%%%%%%%%%%%%%%%%%%%%%%%%%%%%%%%%%%%%%%%%%%%%%%%%%%%%%%%%%%%%%%%%%%%%%%%%%%%%%%%%%%%%%%%%
%%%%%%%%%%%%%%%%%%%%%%%%%%%%%%%%%%%%%%%%%%%%%%%%%%%%%%%%%%%%%%%%%%%%%%%%%%%%%%%%%%%%%%%%%%
%%%%%%%%%%%%%%%%%%%%%%%%%%%%%%%%%%%%%%%%%%%%%%%%%%%%%%%%%%%%%%%%%%%%%%%%%%%%%%%%%%%%%%%%%%

%%%%%%%%%%%%%%%%%%%%%%%%%%%%%%%%%%%%%%%%%%%%%%%%%%%%%%%%%%%%%%%%%%%%%%%%%%%%%%%%%%%%%%%%%%
% Title, Author and Affiliation
%\title{Galactic Radio Signal from the Milky Way Dark Photon Dark Matter Halo}
%\title{Galactic Radio Signal from the Milky Way Electron Density and the Dark Matter Halo}
\title{Dark Photon Dark Matter Radio Signal from the Milky Way Electron Density}

%%%%%%%%%%%%%%%%%%%%%%%%%%%%%%
\author{Ariel Arza\orcid{0000-0002-2254-7408}}
\email{ariel.arza@gmail.com}
\affiliation{Department of Physics and Institute of Theoretical Physics, Nanjing Normal University, Nanjing 210023, China}
%\affiliation{Tsung-Dao Lee Institute, Shanghai Jiao Tong University, Shanghai 200240, China}

%%%%%%%%%%%%%%%%%%%%%%%%%%%%%%%%%%%%%%%%%%%%%%%%%%%%%%%%%%%%%%%%%%%%%%%%%%%%%%%%%%%%%%%%%%
% Abstract
\begin{abstract}
%%%%%%%%%%%%%%%%%%%%%%%%%%%%%
We consider Thomson-like processes between dark matter dark photons and free electrons in the Milky Way. The result is a radio signal background that can be detectable with current or future radio telescope arrays. In particular, we computed sensitivity prospects for the Atacama Large Millimeter Array (ALMA) radio telescope and for the future Square Kilometer Array (SKA), concluding that unconstrained parameter space in a wide range of dark photon masses can be probed.
%%%%%%%%%%%%%%%%%%%%%%%%%%%%%
\end{abstract}
%%%%%%%%%%%%%%%%%%%%%%%%%%%%%%%%%%%%%%%%%%%%%%%%%%%%%%%%%%%%%%%%%%%%%%%%%%%%%%%%%%%%%%%%%%

%%%%%%%%%%%%%%%%%%%%%%%%%%%%%%%%%%%%%%%%%%%%%%%%%%%%%%%%%%%%%%%%%%%%%%%%%%%%%%%%%%%%%%%%%%
\maketitle

The dark matter of the universe is one of the most intriguing problems in fundamental physics. Although its existence is endorsed by astrophysical and cosmological observations, its compositions is a mystery. From the particle physics point of view, several candidates have been proposed, with masses ranging from $10^{-22}\,\text{eV}$ to $10^{19}\,\text{GeV}$ \cite{Bertone:2010zza}.

In the last decade, sub-eV bosons have gained a major interest from the theoretical and experimental community. Such is the case of axions and axion like particles, whose interest has allowed to develop a sophisticated experimental program \cite{Irastorza:2018dyq}. Another well motivated dark matter candidate in the sub-eV mass range is the dark photon \cite{Nelson:2011sf,Arias:2012az,Graham:2015rva,Bastero-Gil:2018uel,Co:2018lka,Dror:2018pdh,Agrawal:2018vin}. Dark photons that are kinetically mixed with standard model photons share many phenomenological aspects with axions, then they can be probed with many experiments designed for axion searches \cite{Caputo:2021eaa,Jaeckel:2012mjv}.

These dark matter candidates can also be searched for with astronomical observations. Axion decays into visible, infrared and radio photons have already been searched from data of real observations \cite{Blout:2000uc,Todarello:2023hdk,Grin:2006aw,Yin:2024lla,Foster:2020pgt}. Furthermore, it was also proposed to look at stimulated decay of axions using radio astronomy equipment \cite{Arza:2019nta,Caputo:2018vmy,Arza:2021nec,Arza:2023rcs,Todarello:2023xuf,Dev:2023ijb,Gong:2023ilg,Buen-Abad:2021qvj,Sun:2021oqp,Ghosh:2020hgd}. For dark photons, astronomical observations have been used on the one hand to study their imprints in the cosmic microwave background \cite{Mirizzi:2009iz,McDermott:2019lch,Caputo:2020bdy} and on the other hand by studying their effects in radio antennas \cite{An:2022hhb,Hardy:2022ufh,An:2023wij,An:2023mvf}.

In this article we propose another mechanism to search for sub-eV dark matter that also requires radio astronomy observations. We focus on radio photons emitted from the interaction between the Milky Way dark matter halo and its free electron density. In particular, we assume the dark matter to be composed entirely by dark photons ($\gamma'$) and focus on the Thomson(Compton)-like scattering process $e^-+\gamma'\to e^-+\gamma$. Since the energy of the photons produced from this process depends strongly on the unknown dark photon mass, the emitted signal may be situated in much of the frequencies of the electromagnetic spectrum. Nevertheless, this work is devoted exclusively to radio emission since higher frequencies require higher dark photon masses, which are already strongly constrained by experiments and astrophysical observations. For very high photon frequencies, the inverse Compton scattering between dark photon dark matter and cosmic rays was studied in Ref. \cite{Su:2021jvk}.

The dark photon interacts with the standard model through a kinetic mixing term with the visible photon\footnote{Other models, where the dark photon interact with the standard model photon through an axion portal, have been proposed. See for instance Refs. \cite{Arias:2020tzl,Deniverville:2020rbv,Hook:2021ous,Domcke:2021yuz,Gutierrez:2021gol,Jodlowski:2023sbi,Hook:2023smg,Hong:2023fcy,DiazSaez:2024dzx}.}. Denoting the visible photon gauge field as $A_\mu$ and the dark photon field as $A_\mu'$, we write the Lagrangian of the theory as
\begin{align}
{\cal L}=&-{1\over4}F_{\mu\nu}F^{\mu\nu}-{1\over4}F_{\mu\nu}'F'^{\mu\nu}+{\mdp^2\over2}A'_\mu A'^\mu \nonumber
\\
&+{\chi\over2}F_{\mu\nu}F'^{\mu\nu}-A_\mu J^\mu, \label{eq:lag}
\end{align}
where $F_{\mu\nu}=\partial_\mu A_\nu-\partial_\nu A_\mu$, $F'_{\mu\nu}=\partial_\mu A'_\nu-\partial_\nu A'_\mu$, $\mdp$ is the dark photon mass, $\chi$ the kinetic mixing parameter and $J_\mu$ the electron current density. The dark photon dark matter field is a zero momentum solution of the system in the propagation eigenstate basis. To get into this basis, we perform the photon field redefinition $A_\mu\to A_\mu+\chi A'_\mu$. In this basis the kinetic mixing term is cancelled and all interactions are manifested in the interaction Lagrangian
\begin{equation}
{\cal L}_I=-A_\mu J^\mu-\chi A'_\mu J^\mu, \label{eq:intlag}
\end{equation}
where it can be noticed that electrons not only have an electric charge $e$ but also a ``dark" charge $\chi e$.

We work in the electron rest frame since the electrons are mostly co-moving with the galactic disk. The dark photon has a non relativistic velocity and dispersion given by $v_{\gamma'}\sim\delta v_{\gamma'}\sim10^{-3}$. The emitted photon from $e^-+\gamma'\to e^-+\gamma$ has an energy $\omega=\mdp\left(1+{\cal O}(v_{\gamma'}^2)\right)$ and can be produced at any direction with almost the same probability. We expect at any location of the galaxy a diffuse spectral line signal background with frequency
\begin{equation}
\nu={\mdp\over2\pi} \label{eq:freq1}
\end{equation}
and bandwidth $\delta\nu\sim\delta v_e\,\nu$, where $\delta v_e$ is the electron velocity dispersion. The electron velocity dispersion can be found through $\delta v_e=\sqrt{3T_e/m_e}$, where $T_e$ is the electron temperature and $m_e$ the electron mass. For simplicity we ignore the small gradient of electron temperature and we assume a nominal value of $T_e=5000\,\text{K}$ \cite{Quireza:2006sn,Dev:2023ijb,Caputo:2018vmy}. Thus, the signal bandwidth is

\begin{equation}
\delta\nu\sim10^{-3}\nu. \label{eq:band1}
\end{equation}

\begin{figure}[t]
\includegraphics[width=1\linewidth]{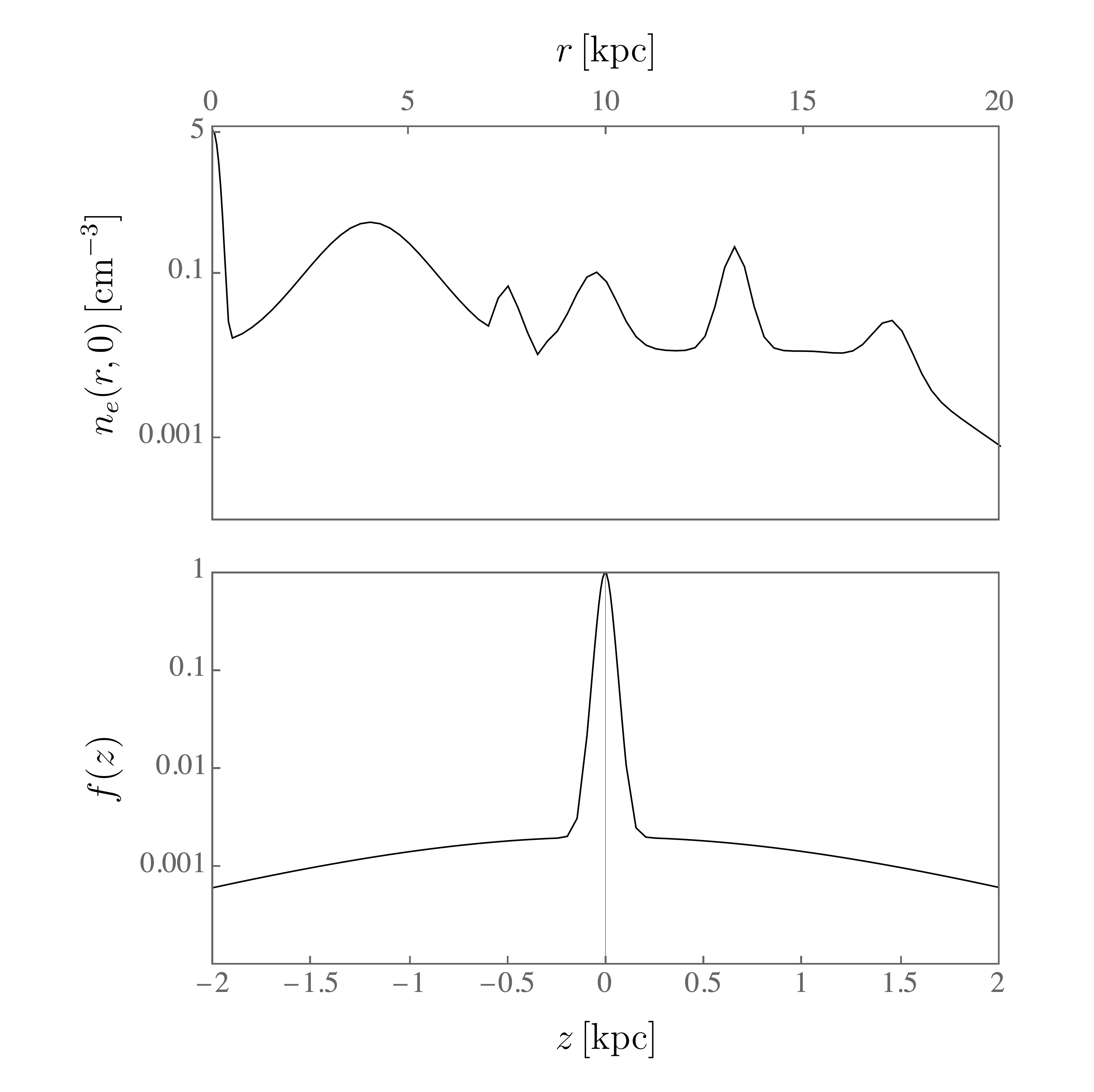}
\caption{Electron density profile in the Milky Way. The top plot shows the profile in the $z$ direction for $r=0$ while the bottom plot shows the profile as a function of $r$ at $z=0$.}
\label{fig:neprof}
\end{figure}

The produced photon energy density $\rho_\gamma$ obeys the Boltzmann equation
\begin{equation}
\dot \rho_\gamma=\left<\sigma v_\text{rel}\right>n_e\,\rhodm, \label{eq:Boltzmann}
\end{equation}
where $n_e$ is the electron density, $\rhodm$ is the dark matter energy density and $\left<\sigma v_\text{rel}\right>$ is the cross section of the process times the relative velocity between the dark photon and the electron, averaged over the corresponding velocity distributions.

In this non relativistic regime, the value of $\sigma v_\text{rel}$ does not depend on the velocities (at least at leading order), then the averaging does nothing. For dark photons we find
\begin{equation}
\sigma v_\text{rel}={8\pi\over3}{\alpha^2\chi^2\over\me^2}=2.07\times10^{-16}\,\chi^2\,\text{cm}^2\,\text{m}/\text{s},
\end{equation}
where $\alpha$ is the fine structure constant. We could also think of searching for axion dark matter using this approach, however for axions with mass $m_a$, the cross section is unfortunately suppressed by a factor $m_a^2/\me^2$, which makes no chances to perform observations in unconstrained parameter space. For scalar dark matter there is not such a suppression, but the current constraints are very strong, even in the range of masses of our interest, therefore this approach is also not promising.

%To compute the radiated intensity $I_\gamma$ at some particular detector, we use the fact that the time derivative of the photon energy density $\dot\rho_\gamma$ in Eq. (\ref{eq:Boltzmann}) is the emitted power per infinitesimal volume $dP/dV$. Then the intensity is found by integrating over the line of sight $\ell$ and the solid angle $\Omega$ covered by the detector. We get

In this work, we choose the observations to be done towards the center of the galaxy since the electron and dark matter densities are huge in there. To compute the signal at some particular detector, we use the fact that the time derivative of the photon energy density $\dot\rho_\gamma$ in Eq. (\ref{eq:Boltzmann}) is the emitted power per unit volume $dP/dV$. The power emitted from an infinitesimal volume, $dV=\ell^2d\ell d\Omega$, located at a distance $\ell$ from the detector, and subtending the solid angle $d\Omega$, is $dP=\left<\sigma v\right>n_e\,\rhodm\ell^2d\ell d\Omega$. As the intensity received by the detector is $dP/(4\pi\ell^2)$, we find that the flux density (power per unit surface per unit frequency) at the detection point is
\begin{equation}
S_\nu={\left<\sigma v\right>\over4\pi\delta\nu}\eta\,\bar n_e\rhodm\,\ell_T, \label{eq:rad1}
\end{equation}
where $\ell_T$ is the total distance in the galactic center direction that crosses the whole galaxy, $\rhodm=0.3\,\text{GeV}/\text{cm}^3$ the nominal value of the dark matter energy density at the sun location, $\bar n_e$ an averaged value for $n_e$ over $\ell_T$, and $\eta$ a model dependent parameter given by
\begin{equation}
\eta={1\over\bar n_e\rhodm}{1\over\ell_T}\int_0^{\ell_T} d\ell\int_\Omega d\Omega\,n_e(\ell,\Omega)\rhodm(\ell,\Omega), \label{eq:eta}
\end{equation}
where the integration is made over the line of sight of the observation and over the solid angle $\Omega$ covered by the telescope.
Eqs. (\ref{eq:rad1}) and (\ref{eq:eta}) do not take into account photon absorption because it is negligible in the frequency range of our interest \cite{Dev:2023ijb}.

\begin{table}[t]
\centering % used for centering table
\begin{tabular}{|l c ||c c |c|} 
\hline %inserts double horizontal lines
halo model & & $\eta$ & & $S_\nu\left[\text{mJy}\right]$   \\ [0.5ex] % inserts table
%heading
\hline\hline % inserts single horizontal line
NFW & & 0.00579 & & 0.238  \\ % inserting body of the table
\hline
Moore & & 0.0116 & &  0.476\\
\hline
Einasto & & 0.0149 & & 0.61 \\  [1ex] % [1ex] adds vertical space
\hline %inserts single line
\end{tabular}
\caption{Flux density estimation for different halo models.} % title of Table
\label{tab:param} % is used to refer this table in the text
\end{table}

The radius of the Milky Way is about $20\,\text{kpc}$ and the distance from us to the galactic center is about $8.3\,\text{kpc}$, then $\ell_T=28.3\,\text{kpc}$. For $n_e(\ell)$ we used the YMW16 model \cite{Yao:2017kcp} (For other models, see \cite{Cordes:2002wz,Gomez:2001te,Taylor:1993my}). For simplicity, we assume azimuthal symmetry for the electron density, with respect to the galactic center. For the profile we use cylindrical coordinates with the origin in the galactic center. We write $n_e(r,z)=f(z)\,n_e(r,0)$, where $z=0$ is attributed to the sun position and $f(z)$ is a decreasing factor that takes into account the vanishing profile in the $z$ direction. The profiles for $n_e(r,0)$ and $f(z)$ are shown in Fig. \ref{fig:neprof}, where the data was extracted from \cite{nedata} along $\ell_T$. From this model, we get\footnote{According to this result, the particle scattering treatment made in this work may not be suitable for masses below $10^{-5}\,\text{eV}$ because of a dark photon Compton wavelength bigger than the average electron separation. However, as will shown in the results, the sensitivity of this proposal in the kinetic mixing parameter is one or two orders of magnitude weaker than current constraints for dark photon masses below $6\times10^{-6}\,\text{eV}$. Therefore there is a small window (in the range $6-10\times10^{-6}\,\text{eV}$) where our proposal is effective that requires a different treatment for signal calculation. Because of this significantly small window compared to the whole space of parameter that we are considering, we leave this issue for future work.} 
\begin{equation}
\bar n_e=0.144\,\text{cm}^{-3}. \label{eq:ne} 
\end{equation}

For $\rhodm(\ell)$ we assumed three different spherically symmetric halo models; the Navarro, Frenk and White (NFW), Moore, and Einasto \cite{Cirelli:2010xx}, which are the most accepted dark matter profiles motivated from N-body simulations. In Tab. \ref{tab:param} we show values of $\eta$ and $S_\nu$ for the three halo models we have chosen. To get $S_\nu$ we have taken $\chi=10^{-11}$, a solid angle of $\Omega=0.26\,\text{deg}^2$, and a signal frequency of $\nu=2.42\,\text{GHz}$ ($\mdp=10^{-5}\,\text{eV}$).

The power signal collected by a single radio telescope with area $A_i$ is
\begin{equation}
P_i=\delta\nu S_\nu A_i, \label{eq:pow1}
\end{equation}
where $S_\nu$ is obtained after integrating over the field of view of the telescope, which is calculated through the formula
\begin{equation}
\fov=\Omega=2\pi(1-\cos(\theta/2)) \label{eq:fov1}
\end{equation}
with $\theta$ given by
\begin{equation}
\theta=1.22{\lambda\over D}, \label{eq:theta}
\end{equation}
where $\lambda$ is the wavelength and $D$ the dish diameter.

The signal to noise ratio for a single dish observation is given by
\begin{equation}
(s/n)_i={P_i\over T_\text{sys}}\sqrt{t_\text{obs}\over\delta\nu}, \label{eq:snr1}
\end{equation}
where $T_\text{sys}$ is the system noise temperature and $t_\text{obs}$ the observation time. Here we have also assumed that the detection system is capable to resolve the small bandwidth $\delta\nu$. For an array composed by $N$ identical dishes, the signal to noise ratio in the single dish mode of the array is
\begin{equation}
(s/n)_\text{array}^\text{sin}=\sqrt{N n_\text{p}}(s/n)_i, \label{eq:snr2}
\end{equation}
where $n_\text{p}$ is the number of polarizations, which in this work is set as $n_\text{p}=2$. For the interferometer mode of the array, the signal in each pixel is determined after integration of Eq. (\ref{eq:eta}) over the synthesized beam. Assuming an homogeneous signal over the entire primary beam, the signal to noise is
\begin{equation}
(s/n)_\text{array}^\text{int}=\sqrt{{N(N-1) n_\text{p}n_\text{pix}\over2}}(s/n)_\text{pix}, \label{eq:snr3}
\end{equation}
where $n_\text{pix}$ is the number of pixels, which is roughly the number of synthesized beams contained in the primary beam.

We can see that for big $N$ the signal to noise is enhanced by a factor $\sqrt{2N}$ for single dish mode operation while for interferometric mode the enhancement goes as $\sim N$. According to these, it seems naively that interferometric mode is more advantageous than the single dish mode. However this is actually not clear because the interferometric mode usually covers a much smaller solid angle in each pixel than the single dish mode. It might lead to a big lose of signal if the source has a big extent, such as the case we are considering in this work.

We calculate sensitivity prospects assuming observations with the currently running ALMA observatory that operates over frequencies ranging from 35 to 950 GHz ($1.45\times10^{-4}\text{eV}<\mdp<3.93\times10^{-3}\text{eV}$). We also take the future SKA radio telescope in its two configurations; SKA low that covers frequencies between 50 and 350 MHz ($2.07\times10^{-7}\text{eV}<\mdp<1.45\times10^{-6}\text{eV}$) and SKA mid with a 0.35 - 13.8 GHz range ($1.45\times10^{-6}\text{eV}<\mdp<5.70\times10^{-5}\text{eV}$). 

For SKA we found the interferometric mode to be more effective than the single dish mode for masses below $\mdp=4\times10^{-6}\,\text{eV}$. For ALMA we find the single dish mode to be more sensitive than the interferometric mode for the whole frequency range. This is an expected result since the interferometric mode provides small synthesized beams at high frequencies. Then it is not very efficient when searching for extended sources.

%For ALMA, since it is equipped with mobile antennas, it is possible to set compact configurations with a wide largest observable scale. We choose the most compact C43-1 configuration \cite{ALMA}, where the interferometer mode is more sensitive than the single dish mode by one order of magnitude in the kinetic mixing parameter. 

%Radio and microwave emission from the center of the galaxy is an important background noise for the planned observations, however, at the frequencies considered in this work, this noise temperature is negligible with respect to the system temperature of ALMA and SKA observations.

\begin{figure}[t]
\includegraphics[width=1\linewidth]{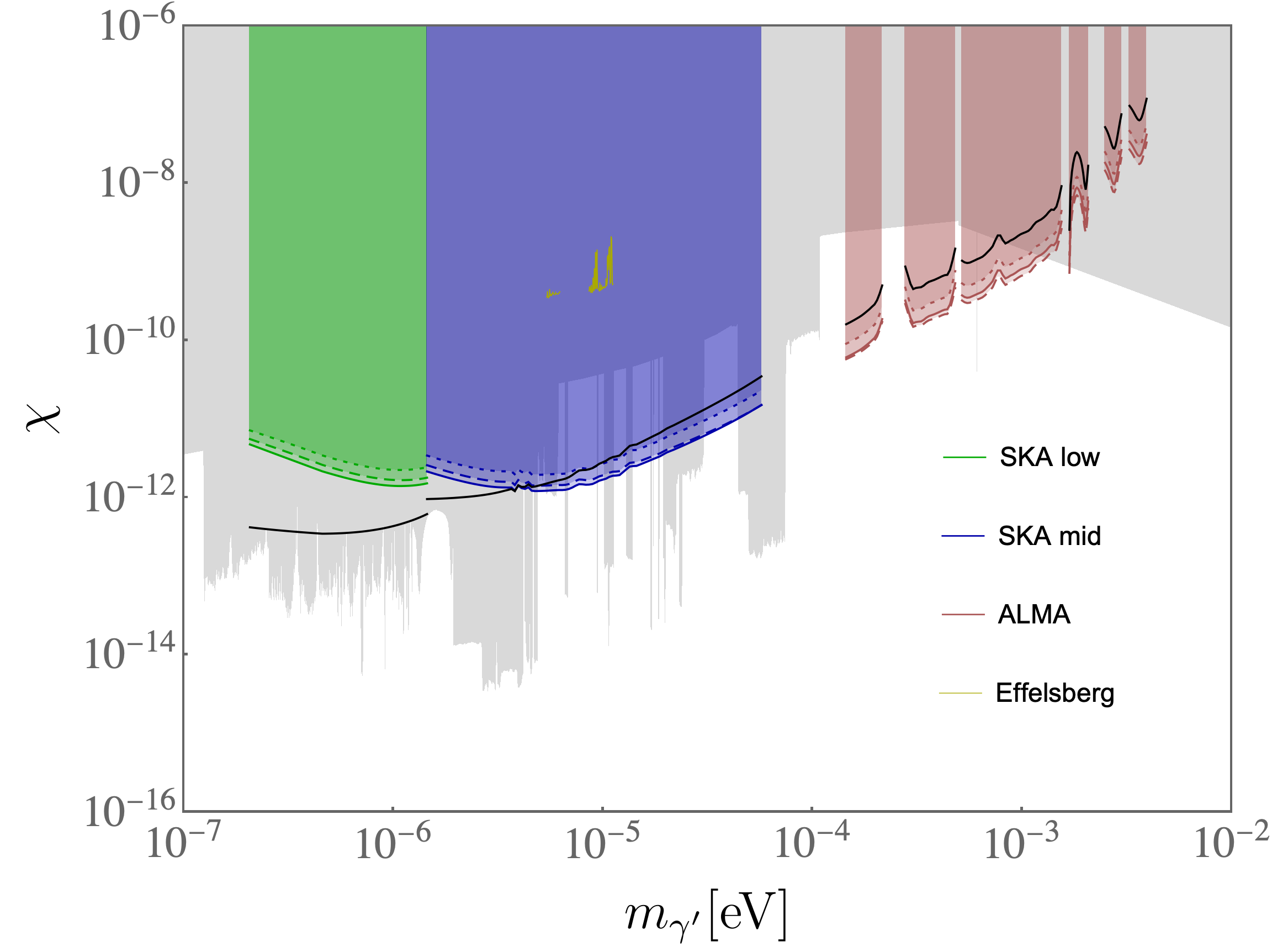}
\caption{Projected sensitivity for 42 days observation with ALMA (red), SKA low (green) and SKA mid (blue) in single dish operation. The dotted, dashed and solid lines correspond to NFW, Moore, and Einasto halo models, respectively. The black solid lines correspond to the projected sensitivities in the interferometric mode assuming the Einasto profile. The yellow line shows constraints from flux density limits obtained from real observations of the Effelsberg telescope described in Ref. \cite{Foster:2020pgt}. The grey region are current experimental, astrophysical and cosmological constraints}
\label{fig:sens}
\end{figure}

Our results are shown in Fig. \ref{fig:sens}, where we have assumed a total observation time of $t_\text{obs}\approx42\,\text{days}$ and a signal to noise ratio of $s/n=5$. The colored regions correspond to sensitivity projections for ALMA (red) SKA mid (blue) and SKA low (green) in the single dish mode. The projections in the interferometer mode are marked with a black solid line.

For ALMA we considered their 54 dishes with 12 m diameter and their 12 dishes with 7 m diameter \cite{ALMA}. For the single dish mode of the array, we compute the solid angle covered by each antenna using Eq. (\ref{eq:fov1}) and Eq. (\ref{eq:theta}).
 
For the interferometer mode, the synthesized beam is determined empirically by the formula \cite{ALMA}
\begin{equation}
\theta_\text{synth}=0.574{\lambda\over L_{80}}, \label{eq:sinthal}
\end{equation}
where $L_{80}$ is the $80^\text{th}$ percentile projected baseline. On the other hand, the maximum recoverable scale (MRS) which a given array is sensitive, is determined by
\begin{equation}
\theta_\text{MRS}=0.983{\lambda\over L_{5}}, \label{eq:MRSal}
\end{equation}
where $L_{5}$ is the $5^\text{th}$ percentile projected baseline.

For the ALMA forecast in the interferometric mode, we assume the very compact configuration C43-1 which is convenient for extended sources since it has very short baselines. For C43-1 we have $L_{80}=107\,\text{m}$ and $L_5=21.4,\text{m}$. For this case, the number of pixels is roughly $n_\text{pix}\sim\theta_\text{MRS}^2/\theta_\text{synth}^2\sim73$.

\begin{figure}[t]
\includegraphics[width=1\linewidth]{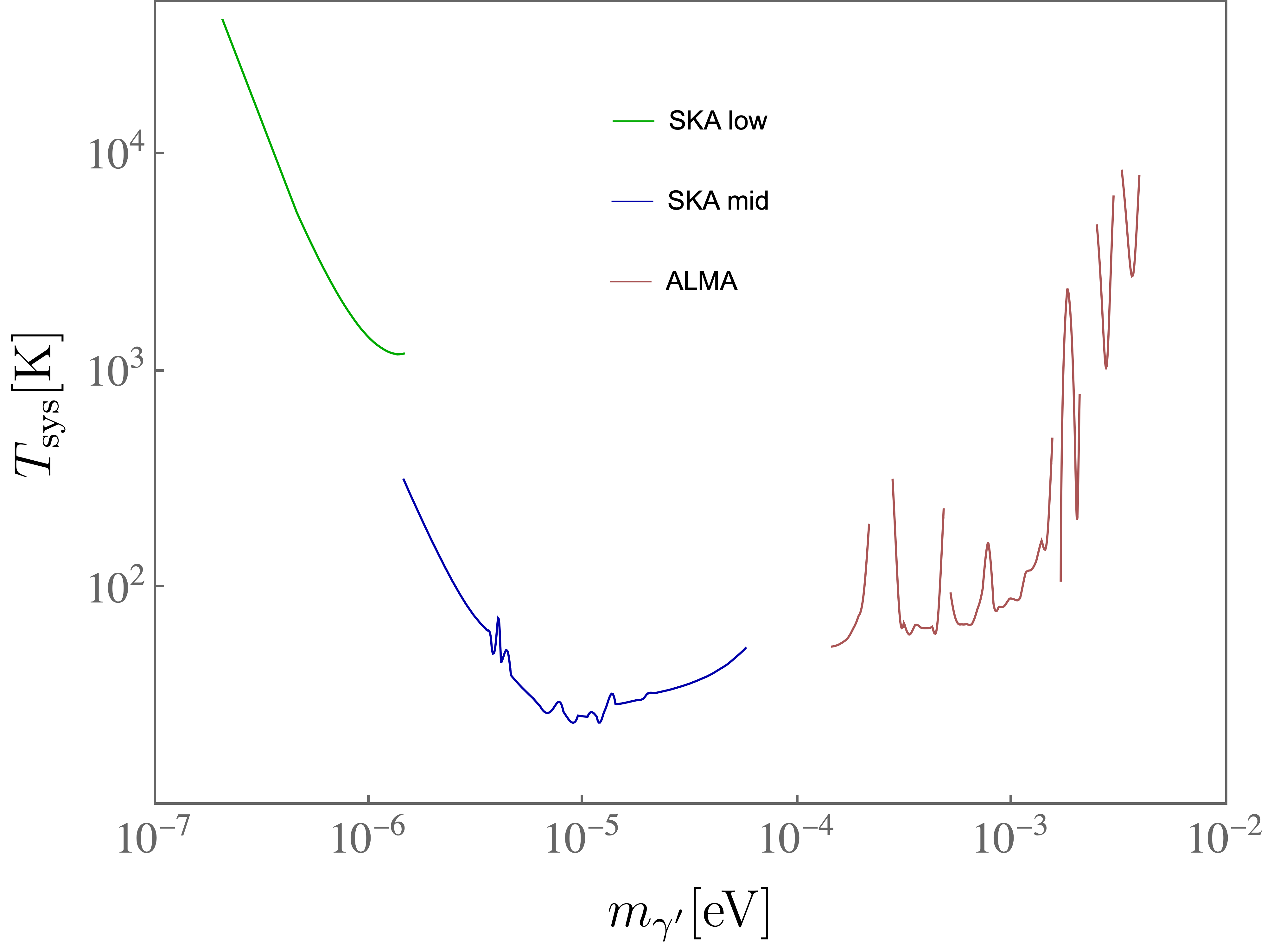}
\caption{System temperature $T_\text{sys}$ for ALMA (red), SKA low (green) and SKA mid (blue).}
\label{fig:Tsys}
\end{figure}

For $T_\text{sys}$ we have collected the values given in the sensitivity calculator of the official ALMA webpage \cite{ALMATsys}. It includes the receiver noise and all the background noise from the sky pointing to the galactic center. The total system temperature for the whole ALMA frequency band is shown in Fig. \ref{fig:Tsys}. Emissions from the galactic center within ALMA frequency range are, for example, reported in \cite{Guan:2021wmh}. They get fluxes not bigger than $70\,\text{kJy}$ from the inner $1.5^\circ\times0.5^\circ$ region. These fluxes correspond to noise temperature values smaller than the ones used in this work to compute the sensitivities. The sensitivity forecast is shown in Fig. \ref{fig:sens}, where the red regions correspond to sensitivity prospects for the six operation bands; 35 - 52 GHz, 67 - 116 GHz, 125 - 373 GHz, 385 - 500 GHz, 602 - 720 GHz and 787 - 950 GHz, in the single dish mode, for the three dark matter profiles considered in this work. The solid black line is our projected sensitivity for the Einasto profile assuming the interferometric mode with the C43-1 configuration.

For SKA low we consider the 512 stations, each with a diameter of 35 m. For SKA mid we take the whole array consisting of 133 dishes with 15 m of diameter and other 64 dishes with a diameter of 13.5 m \cite{SKA}. For the single dish mode of both SKA low and SKA mid, we compute the solid angle covered by each antenna using Eq. (\ref{eq:fov1}) and Eq. (\ref{eq:theta}). For the interferometric mode we assume a synthesized beam equivalent to a longest baseline of 1 km and a primary beam equivalent to a shortest baseline of the same order of the dishes diameter. Thus, the number of pixels is about 816 for SKA low and 4400 for SKA mid.

Concerning the noise temperature, we extracted data for the combined parameter $A/T_\text{n}$ plotted in \cite{SKA} as a function of the frequency, corresponding for the whole array for SKA low and valid for a single dish for SKA mid. From these we calculate $T_n$, which contains the noise temperature of the receiver and the sky. This noise temperature was collected from a direction which does not point to the galactic center. To fix this we first subtract the sky temperature contained in $T_n$, which dominates the low frequency part of both SKA low and SKA mid with a simple power law. Now, to get $T_\text{sys}$ we add the background noise temperature $T_\text{GC}$ coming from the galactic center. For this, we assume the simple model described in \cite{Yusef-Zadeh:2012efm}, valid for frequencies above 350 MHz, and that uses observations from the inner $2^\circ\times0.85^\circ$ of the galaxy. To extend the model to lower frequencies, we used the results obtained in \cite{Guzman:2010da} to get a power law for frequencies between 45 and 408 MHz. Matching both models, we get a background noise temperature model of the form $T_\text{GC}(\nu)=15.3\,\text{K}\,F(\nu)$, where the function $F(\nu)$ is $0.606(\nu/\nu_0)^{-2.517}$ for $\nu<325\,\text{MHz}$, $(\nu/\nu_0)^{-2.173}$ for $\nu<\nu_0$, $(\nu/\nu_0)^{-2.58}$ for $\nu_0<\nu<4.85\,\text{GHz}$ and $1.99(\nu/\nu_0)^{-3.14}$ for $\nu>4.85\,\text{GHz}$. The value of $\nu_0$ is taken as $1.41\,\text{GHz}$. The total system temperature for SKA low and SKA mid are shown in Fig. \ref{fig:Tsys}.

The green region of Fig. \ref{fig:sens} shows the sensitivity forecast for SKA low that will operate in the frequency range 50 - 350 MHz, while the blue region does so for SKA mid that will operate in the range 0.35 - 13.8 GHz. Both colored regions correspond to the single dish mode for the three galactic halo profiles, while the solid black lines is the equivalent sensitivity for the Einasto profile assuming inteferometric mode.

So far, all the sensitivity projections were computed taking into account the noise of the receivers, the background noise around the telescope and also the background noise coming from the galactic center, which is our observation target. One irreducible background we have not discussed yet is the presence of spectral lines whose widths are similar to the ones expected for the dark photon signal. Most of the spectral lines, for instance recombination lines \cite{Brown:1978wv}, are very well known and are easy to identify. The only ones that might be problematic, are the 21 cm lines, which are sufficiently strong for redshifts $z<6$. These correspond to dark photon masses in the range $8\times10^{-7}-6\times 10^{-6}\,\text{eV}$, which are already excluded by axion haloscope experiments (see Fig. \ref{fig:sens}). Then we just ignore 21 cm lines in this work. There might be unlikely cases where known spectral lines overlap with our dark photon signal. However, as this work is in the early stage of this proposal, we think the analysis of how to remove such a background (which may include, for instance, the reconstruction of the morphology of the dark photon line signal) would be appropriate at future work.

Another important point worth to mention is that for the sensitivity projections we have used the ideal Dicke's radiometer equation which assumes that all the noise is thermal. In reality the sensitivities might be slightly weaker because of sources of systematic uncertainties such as is pointed out in \cite{Foster:2020pgt}. This reference provides real sensitivity limits on the galactic center flux density for observations performed by the Effelsberg telescope with an integration time of 61.9 and 40 minutes in the S-band and L-band, respectively. We take this flux density limits and compare directly to Eq. (\ref{eq:rad1}) (assuming an Einasto profile) to put constraints on dark photon dark matter. This constraint is shown with a yellow solid line in a small region of Fig. \ref{fig:sens}, between 5.29 and 11.2 $\mu\text{eV}$. This constraint is not competitive with the ones from current haloscope experiments. Of course, the sensitivity prospects from SKA are much stronger because of the much better capabilities that SKA has.

In this work we have proposed to search for hypothetical dark photon dark matter by detecting the radio signal produced by Thomson-like scattering with galactic free electrons. We computed sensitivity forecasts for radio telescope arrays such as ALMA and SKA, and concluded that this search is promising to explore dark photon dark matter in unconstrained parameter space. ALMA has sensitivity for masses that are not accessible with axion haloscopes, while SKA can be used as a complementary search to them, that also has the potential to fill their gaps in parameter space.

%Despite the sensitivity forecast was done using realistic data, details on the uncertainties of the projected sensitivities as well as background subtraction  from other radio sources are beyond the scope of this work. We think it is appropriate to address these issues in future work by using real data from observations.  

We would like to thank Bin Zhu for his advice on numerical integration and also Gwenael Giacinti, Nick Houston, Marco Regis, Elisa Todarello, Lei Wu and Xiaolong Yang for useful discussions.

\bibliographystyle{JHEP}
\bibliography{references.bib}

\end{document}